\documentclass[10pt,letterpaper]{article}
\usepackage{opex3}
\usepackage{graphicx}
\usepackage{overpic}
\usepackage{subfig}

\begin{document}
\title{Genetic Optimization of Photonic Bandgap Structures}
\author{Joel Goh, Ilya Fushman, Dirk Englund, Jelena Vu\v{c}kovi\'{c}}
\address{Ginzton Laboratory, Stanford University, Stanford, CA 94305, USA}
\email{joelgoh@stanfordalumni.org; ifushman@stanford.edu; englund@stanford.edu;
jela@stanford.edu}

\begin{abstract}
We investigate the use of a Genetic Algorithm (GA) to design a set of photonic crystals
(PCs) in one and two dimensions. Our flexible design methodology allows us to optimize
PC structures which are optimized for specific objectives. In this paper, we report the
results of several such GA-based PC optimizations. We show that the GA performs well
even in very complex design spaces, and therefore has great potential for use as a
robust design tool in present and future applications.
\end{abstract}




\section{Introduction}
Photonic crystals (PCs) describe a class of semiconductor structures which exhibit a
periodic variation of refractive index in 1, 2, or 3 dimensions. As a result of this
periodic variation, PCs possess a photonic band gap -- a range of frequencies in which
the propagation of light is forbidden~\cite{John1, Yab1}. This is the analog of the
electronic bandgap in traditional semiconductors. This unique characteristic of PCs
enables them to be used to effectively manipulate light. PCs have already been used for
applications such as modifying the spontaneous emission rate of emitters ~\cite{Eng1,
Bor1}, slowing down the group velocity of light~\cite{Alt1, Vla1}, and designing highly
efficient nanoscale lasers~\cite{Alt2}.

Given that Photonic Crystals find applications in a myriad of areas, we proceed to
investigate the question: \emph{What is the best possible PC design for a given
application?} Traditionally, the design of optimal PC structures has been largely done
by either trial-and-error, iterative searches through a design space, by physical
intuition, or some combination of the above methods~\cite{Song1, Vuc1}. However, such
methods of design have their limitations, and recent developments in PC design
optimization have instead taken on a more systematic and algorithmic nature ~\cite{Eng2,
Jia1, Pre1, Drup1}. In this work, we report the results of a Genetic Algorithm to
optimize the design of a set of one and two-dimensional PC structures. We show that the
Genetic Algorithm can effectively optimize PC structures for any given design objective,
and is thus a highly robust and useful design tool.


\section{Genetic Algorithms}
Genetic Algorithms (also known as Evolutionary Algorithms) are a class of optimization
algorithms that apply principles of natural evolution to optimize a given
objective~\cite{Hol1, Gol1, Dav1}. In the genetic optimization of a problem, different
solutions to the problem are picked (usually randomly), and a measure of fitness is
assigned to each solution. On a given generation of the design, a set of operations,
analogous to mutation and reproduction in natural selection, are performed on these
solutions to create a new generation of solutions, which should theoretically be
``fitter'' than their parents. This process is repeated until the algorithm terminates,
typically after a pre-defined number of generations, or after a particularly ``fit''
solution is found, or more generally, when a generation of solutions meets some
pre-defined convergence criterion.


\section{Implementation}
Genetic Algorithms have already been used in PC design - to find non-intuitive
large-bandgap designs~\cite{Pre1, Shen1} and for designing PC fibers~\cite{Ker1}. In our
work, we performed the genetic optimization by varying the sizes of circular holes in a
triangular lattice. This approach was chosen because the search space is conveniently
constrained in this paradigm, and the optimized structures can be easily fabricated, if
desired. A freely available software package~\cite{Johnson1} was used to simulate the
designed structures.

In addition, we used the following parameters for the implementation of our Genetic
Algorithm:

\medskip
\begin{description}
\item{\textbf{Chromosome Encoding.}} We used a direct-chromosome encoding, where the various optimization
parameters were stored in a vector. For the current simulations, for simplicity, we only
varied the radii of cylindrical holes in a triangular lattice. Our implementation can be
easily modified to include other optimization parameters as well, such as the positions
of the various holes, or the refractive index of the dielectric material.

\smallskip
\item{\textbf{Selection.}} We used fitness-proportionate selection (also known as roulette-wheel
selection), to choose parent chromosomes for mating. In this selection scheme, a
chromosome is selected with a probability $P_i$ that is proportional to its fitness
$f_i$, as shown in Eq. (\ref{roulette}).

\begin{equation}\label{roulette}
    P_i = \frac{f_i}{\displaystyle\sum_{k=1}^{N}f_k}
\end{equation}

\smallskip
\item{\textbf{Mating.}} After a pair of parent chromosomes $v_{parent,1}$ and $v_{parent,2}$
were selected, they were mated to produce a child chromosome $v_{child}$ by taking a
random convex combination of the parent vectors, as in Eq. (\ref{mating}).
\begin{eqnarray}\label{mating}
    \nonumber \lambda &\sim& U(0, 1) \\
    \vec{v}_{child} &=& \lambda \vec{v}_{parent,1} + (1-\lambda)\vec{v}_{parent,2}
\end{eqnarray}

\smallskip
\item{\textbf{Mutation.}} Mutation was used to introduce diversity in the population. We
used two types of mutation in our simulations, a random-point crossover and a gaussian
mutation.

\textbf{1) Random-point crossover}: For an original chromosome vector $\vec{v}_{orig}$
of length $N$, we select a random index, $k$, from 0 to $N$ as the crossover point, and
swap the two halves of $\vec{v}_{orig}$ to produce the mutated vector, $\vec{v}_{mut}$,
as represented in Eq. (\ref{mutation}).
\begin{eqnarray}
    \nonumber   \vec{v}_{orig} &=& (v_1, v_2, \ldots, v_N)^T \\
    \nonumber    \qquad k &\sim& U\{0, 1, 2, .... , N\} \\
    \label{mutation} \vec{v}_{mut} &=& (v_{k+1}, v_{k+2}, \ldots, v_N,
                                    v_1, v_2, \ldots, v_{k-1})^T
\end{eqnarray}

\textbf{2) Gaussian mutation}: To mutate a chromosome vector by Gaussian mutation, we
define each element of $\vec{v}_{mut}$ to be independent and identically distributed
Gaussian Random Variables with mean $\vec{v}_{orig}$ and a standard deviation
proportional to the corresponding elements of $\vec{v}_{orig}$. This searches the space
in the \emph{vicinity} of the original chromosome vector $\vec{v}_{orig}$.
\begin{eqnarray}
    \label{gauss_mutation} v_{i}^{mut} \sim N\left(v_{i}^{orig}, \sigma^2 \right),
            \qquad i \in \{0, 1, 2, .... , N\}
\end{eqnarray}

$\sigma^2$ is a algorithm-specific variance, and can be tuned to change the extent of
parameter-space exploration due to mutation.

\smallskip
\item{\textbf{Cloning.}} To ensure that the maximum fitness of the population was would
never decrease, we copied (cloned) the top few chromosomes with the highest fitness in
each generation and inserted them into the next generation.
\end{description}

\section{Simulation Results}
\subsection{Optimizing Planar Photonic Cavity Cavities}
\subsubsection{\emph{Q}-factor Maximization}
One problem of interest in PC design is the inverse problem, where one tries to find a
dielectric structure to confine a given (target) electromagnetic mode.  Here we consider
the inverse design problem of optimizing a linear-defect cavity in a planar photonic
crystal cavity. The \emph{Q}-factor is a common figure of merit measuring how well a
cavity can confine a given mode, and can be approximated (assuming no material
absorption) by the following expression:

\begin{equation}\label{q_factor}
    \frac{1}{Q_{\emph{total}}} = \frac{1}{Q_{||}} + \frac{1}{Q_{\bot}}
\end{equation}

where $Q_{||}$ represents the \emph{Q}-factor in the direction parallel to the slab, and
$Q_{\bot}$ represents the \emph{Q}-factor perpendicular to the slab. $Q_{\bot}$ is
usually the limiting factor for $Q_{\emph{total}}$. As was shown previously~\cite{Eng2,
Aka1}, the vertical mode confinement, which occurs through total internal reflection
(TIR), can be improved if the mode has minimal k-space components inside the light cone.

In the subsequent sections, we report the results where we employed our GA to minimize
the light cone radiation of such cavities. We used one-dimensional photonic crystals as
approximations to these cavities~\cite{Lal1}, and simulated these cavities using the
standard Transfer Matrix method for the E-field~\cite{Yar1}.

\subsubsection{Matching to a Target Function}
In~\cite{Eng2} it was noted that minimization of light cone radiation could be performed
via mode-matching to a target function which already possessed such a property. We
therefore used a fitness function that was equal (up to a normalizing factor) to the
reciprocal of the mean-squared difference between our simulated mode and a target mode
(see Eq~(\ref{fitness_1D})). For this simulation, our chromosome encoded the thicknesses
of the dielectric slabs in the structure, and was a vector of length 10. We used 100
chromosomes in each generation and allowed them to evolve for 80 generations.

\begin{equation}\label{fitness_1D}
    fitness \propto \left\{\int_{-\infty}^{\infty}{|f_{sim}(x) -
f_{target}(x)|^2}dx\right\}^{-1}
\end{equation}

\begin{figure}[htp]
    \centering
    \includegraphics[width=5cm]{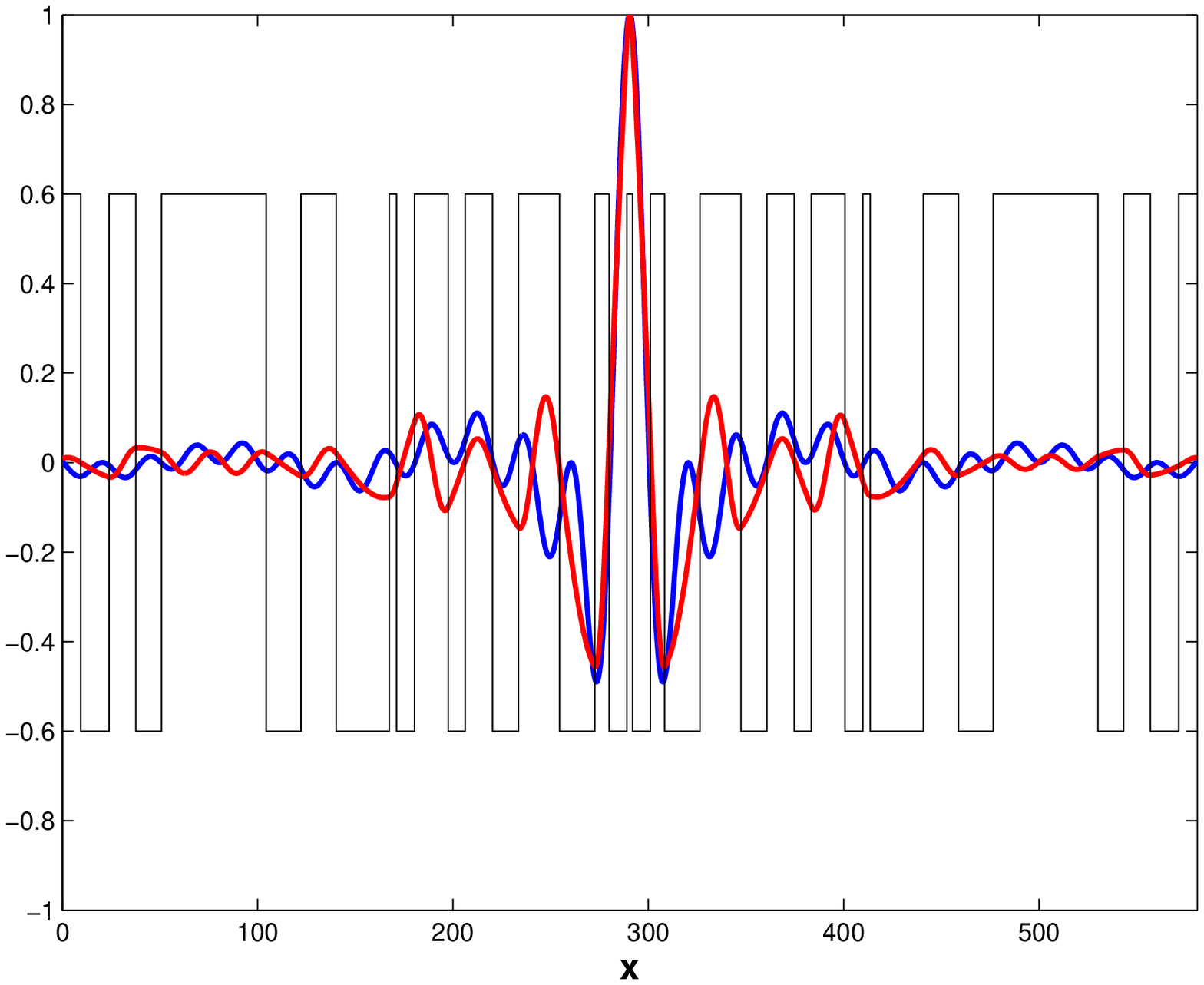}
    \includegraphics[width=5cm]{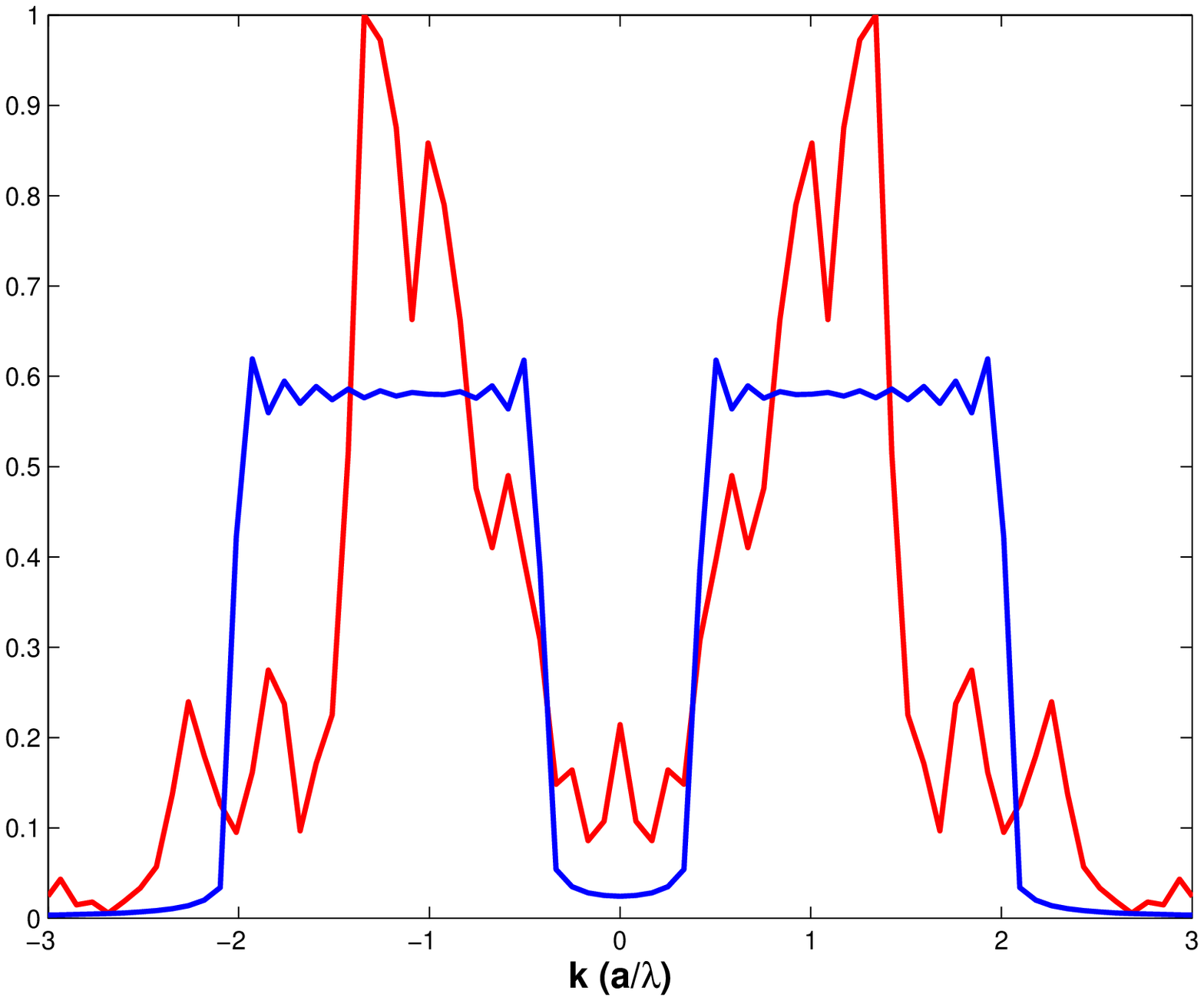}\\
    \includegraphics[width=5cm]{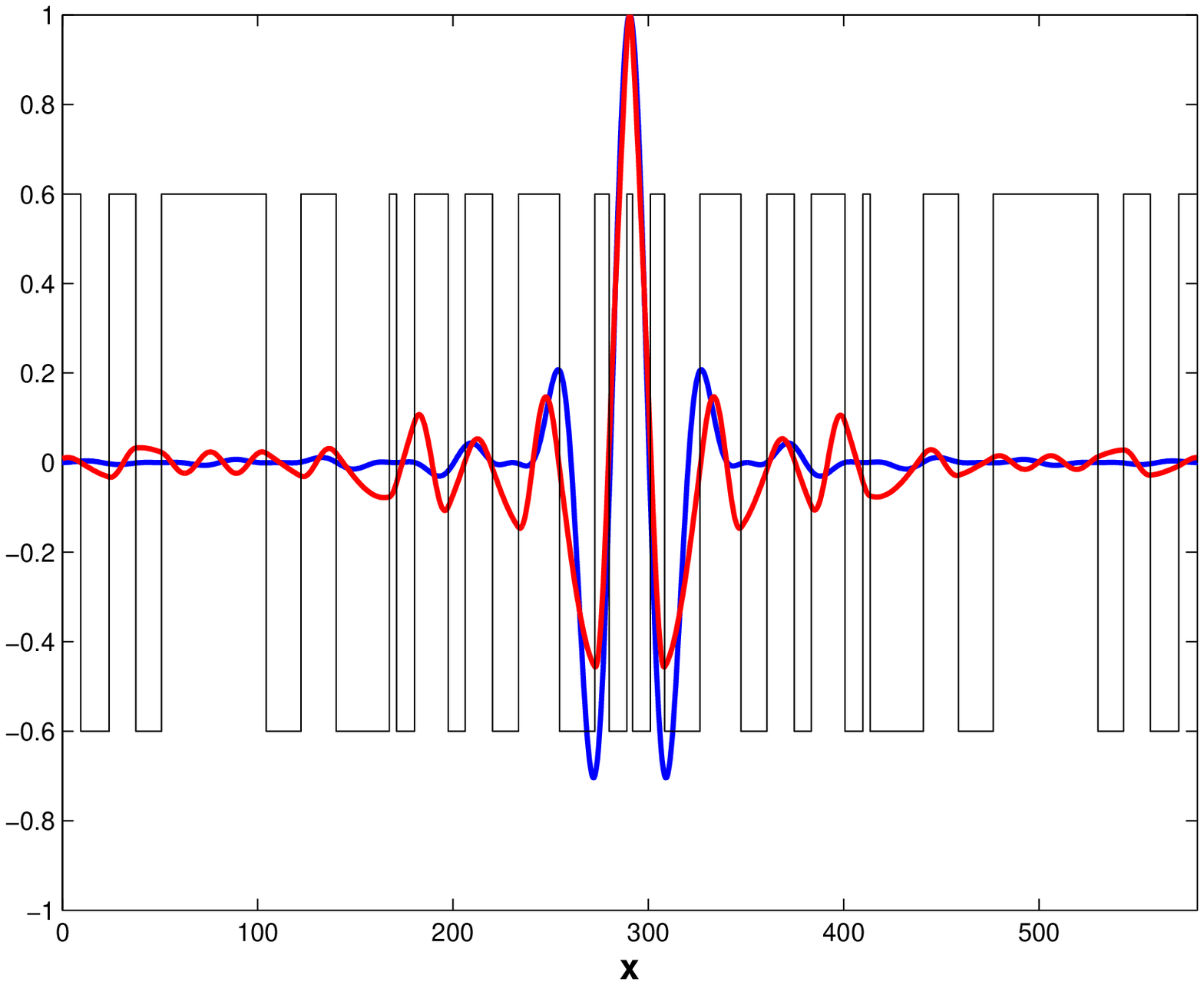}
    \includegraphics[width=5cm]{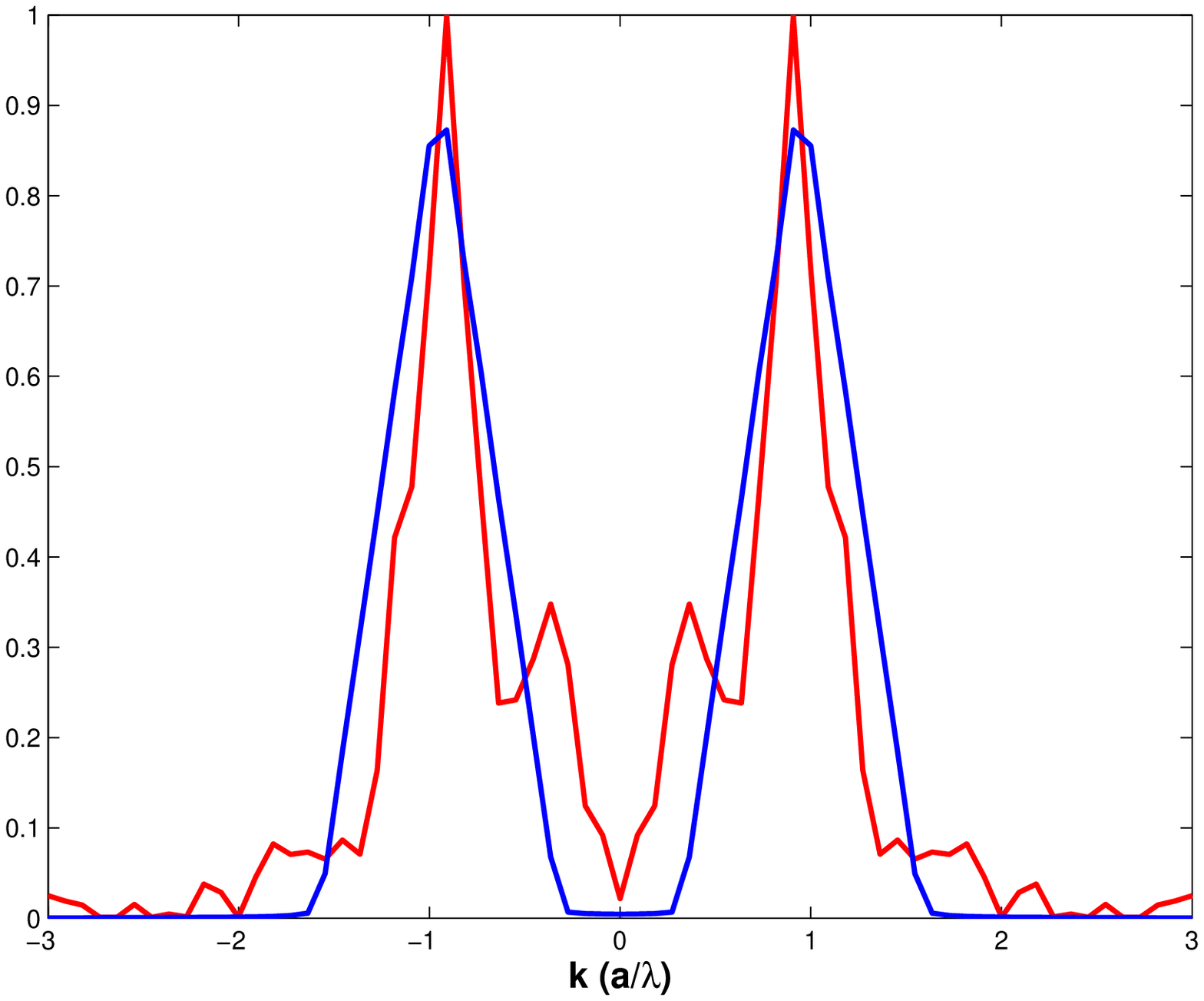}
    \caption{\emph{Top-left}: Real-space mode profile after optimizing for closest-match
    to a sinc-envelope target mode. \emph{Top-right}: k-space mode profile of optimized simulated mode and
    a sinc-envelope target mode. \emph{Bottom}: Real-space and k-space mode profiles for matching
    against a sinc$^2$-envelope target mode.}
    \label{match_to_tgt_fn}
\end{figure}

We used target modes that were sinusoidal functions multiplied by \emph{sinc} and
\emph{sinc-squared} envelope respectively. Such target modes have theoretically no
radiation at or near the Gamma point and are therefore ideal candidates as target
functions. The results, shown in Fig.~\ref{match_to_tgt_fn}, clearly feature a
suppression of k-vector components at low spatial-frequencies. Matching using the the
\emph{sinc-squared} envelope target function appeared to produce a better match. From
the k-space plots, the GA evidently had difficulty matching the sharp edges for the
\emph{sinc}-envelope target mode.

\medskip

\subsubsection{Direct Minimization of Light Cone Radiation}
In the preceding subsection, we observed that when we formulated our objective as a
matching problem, in the case of the \emph{sinc}-envelope, the GA sacrificed the desired
low spatial-frequency suppression in an effort to match the overall shape of the
function. The preceding formulation therefore poses an implicit constraint on our
optimization. By reformulating the optimization problem, we were able to effectively
remove this constraint, and obtain a better result.

Our reformulation directly minimized the k-vector components in the light cone, by
minimizing the integrated square-magnitude of the simulated E-field mode in k-space
inside the light cone. The fitness function that we used is given as in
Eq~(\ref{fitness_direct}), where \emph{V} represents the set of k-vectors within the
light cone.

\begin{equation}\label{fitness_direct}
    fitness = \left\{\int_{V}{|F(k)|^2dk}\right\}^{-1}
\end{equation}

\begin{figure}[htp]
    \centering
    \includegraphics[width=8cm]{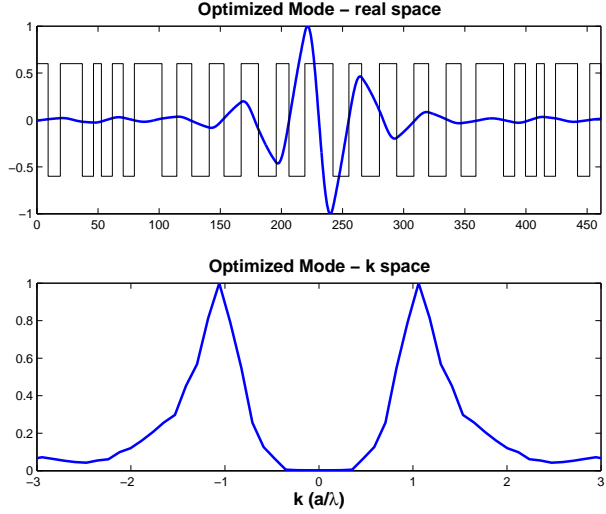}
    \caption{\emph{Top}: Real-space mode profile of optimized resonant E-field mode.
    \emph{Bottom}: Corresponding k-space mode profile of optimized mode}
    \label{min_lc_mode}
\end{figure}

The final, evolved structure, together with the corresponding real-space and k-space
mode profiles are shown in Fig~\ref{min_lc_mode}. The k-space mode profile features a
strong suppression of radiation at low frequencies, to a greater extent as compared to
the optimized fields from the preceding simulations. By relaxing our constraint and
performing a direct optimization, our GA has designed a structure that achieves better
light cone suppression than before. Our direct optimization paradigm has exploited the
extreme generality of the GA, which simply requires that a fitness function be defined,
with little further constraint thereafter.

\medskip

\subsection{Maximal Gap at any k-vector Point}
Moving on to the more generic case of 2D photonic crystals, we will proceed to show the
results of simulations for maximizing the TE bandgap at any point in k-space for a
2-Dimensional PC structure with a triangular lattice of air holes. This could be useful
for PC design applications where the target mode to be confined is centered around a
particular point in k-space~\cite{Eng2}. By maximizing the bandgap at that k-space
point, we would effectively design a better mirror for a mode resonating along this
k-space direction.

We used a supercell which was three periods wide in each dimension and varied the radii
of the nine holes in total, and we encoded the chromosome as a vector of these nine
holes. We used a population size of 60 chromosomes for each generation, and allowed the
optimization to run for a total of 100 generations.

\begin{figure}[htp]
    \centering
    \includegraphics[width=8cm] {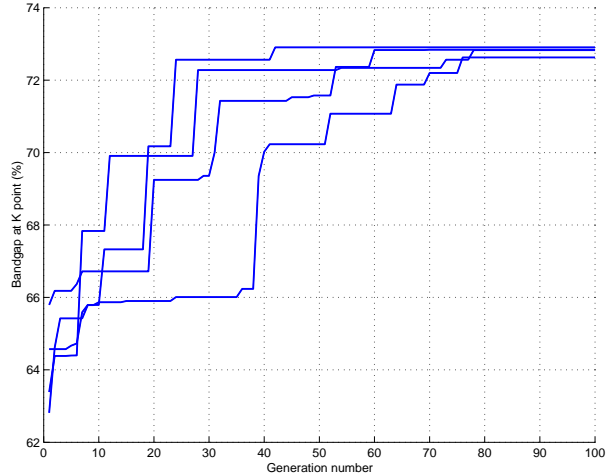}
    \caption{Fitness (gap-to-midgap ratio at K-point of the band diagram) of
    maximally-fit structure of each generation for 100 generations.
    The maximum fitness is a monotonically non-decreasing function due to cloning.
    A general increase in fitness arises as a result of various genetic operations
    (selection, mating, mutation).}
    \label{maxKpoint}
\end{figure}

To evaluate the fitness of each chromosome, we used the eigensolver in
Ref~\cite{Johnson1} to calculate the gap-to-midgap ratio at the K-point of the band
diagram. We then scaled the calculated ratio exponentially to tune the selection
pressure of the optimization. Figure~\ref{maxKpoint} shows the variation of the
gap-to-midgap ratio of our structures as the algorithm progressed.

\begin{figure}[htp]
    \centering
    \subfloat[Run 1]{\label{maxKpoint-dielectric-a}
    {\begin{overpic}[width=2.8cm]{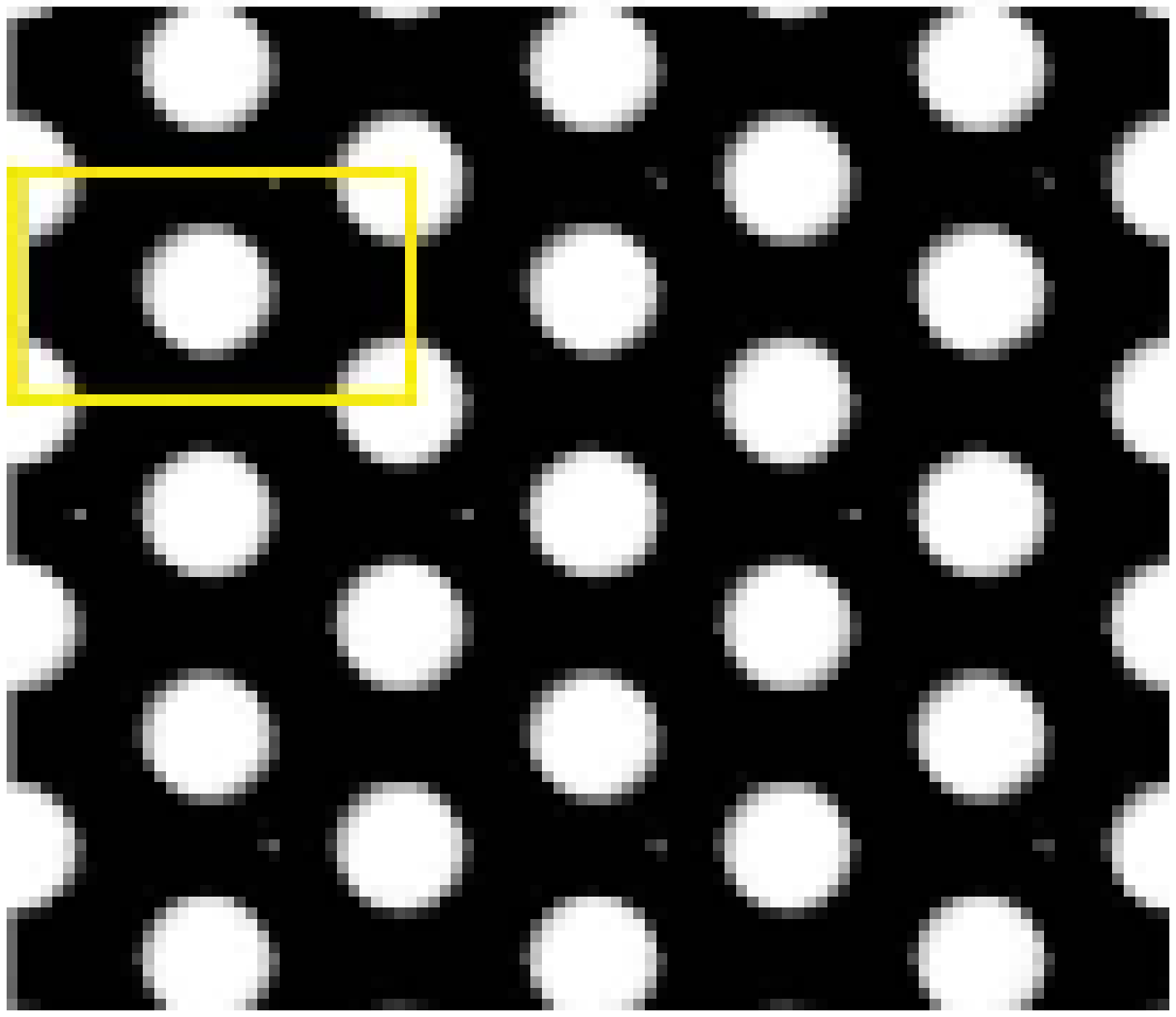}\end{overpic}}}
    \subfloat[Run 2]{\label{maxKpoint-dielectric-b}
    {\begin{overpic}[width=2.8cm]{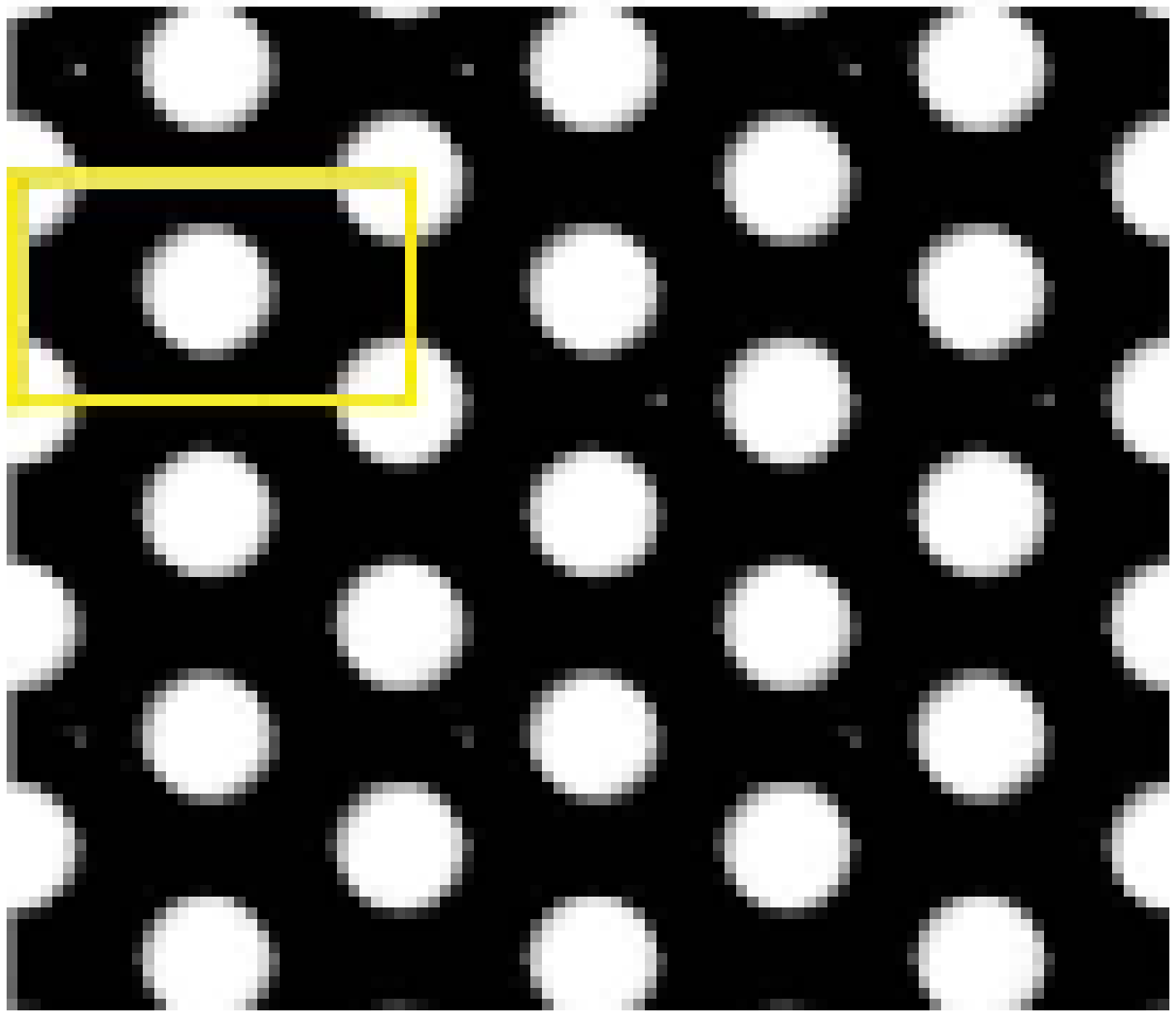}\end{overpic}}}\\
    \subfloat[Run 3]{\label{maxKpoint-dielectric-c}
    {\begin{overpic}[width=2.8cm]{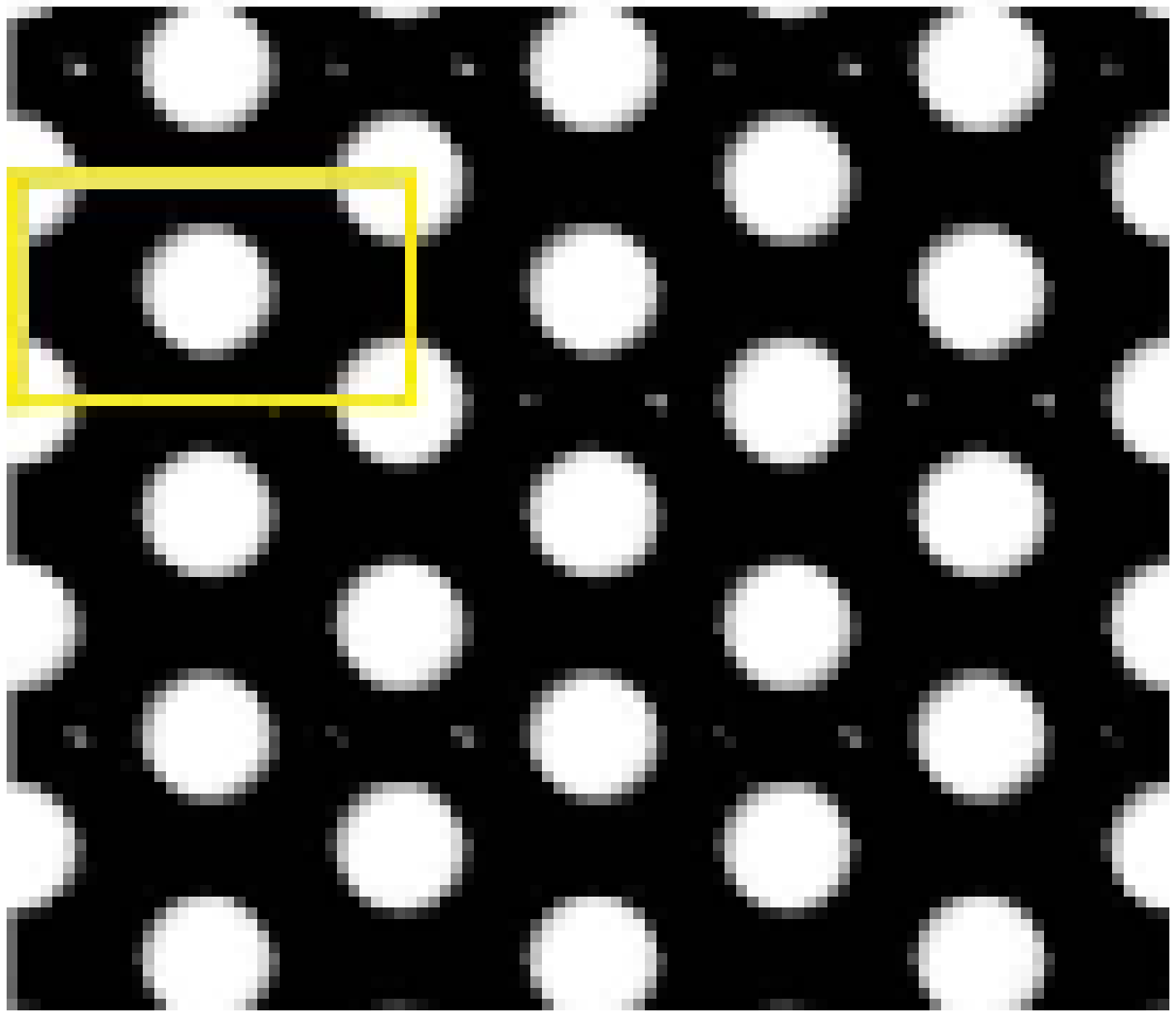}\end{overpic}}}
    \subfloat[Run 4]{\label{maxKpoint-dielectric-d}
    {\begin{overpic}[width=2.8cm]{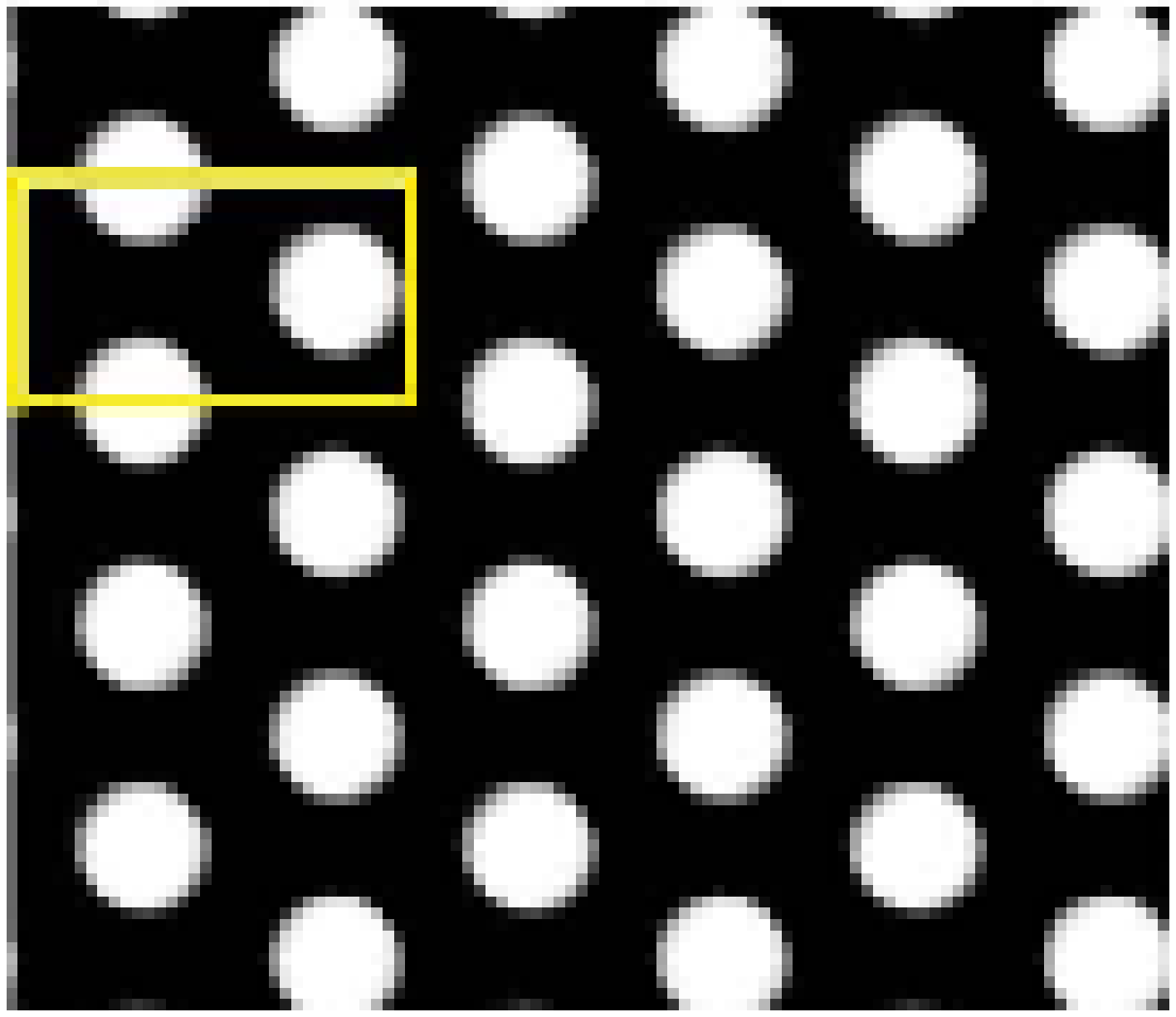}\end{overpic}}}\\
    \subfloat[Band Diagram - optimized]
       {\label{banddiag_run3}
      {\begin{overpic}[width=8cm]{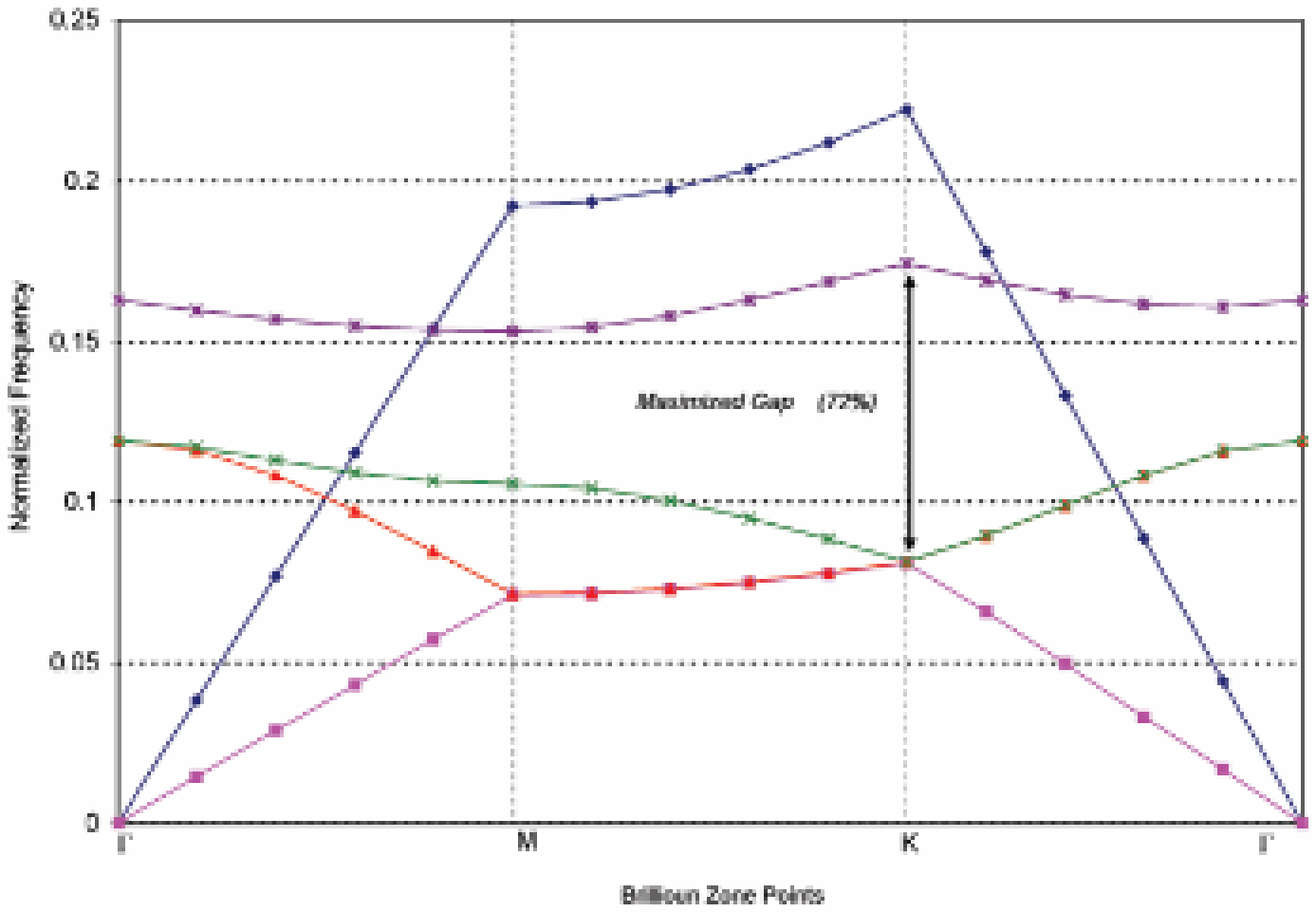}\end{overpic}}}\\
    \subfloat[Band Diagram - uniform holes, r/a = 0.3]
        {\label{banddiag_uniform}
        {\begin{overpic}[width=8cm]{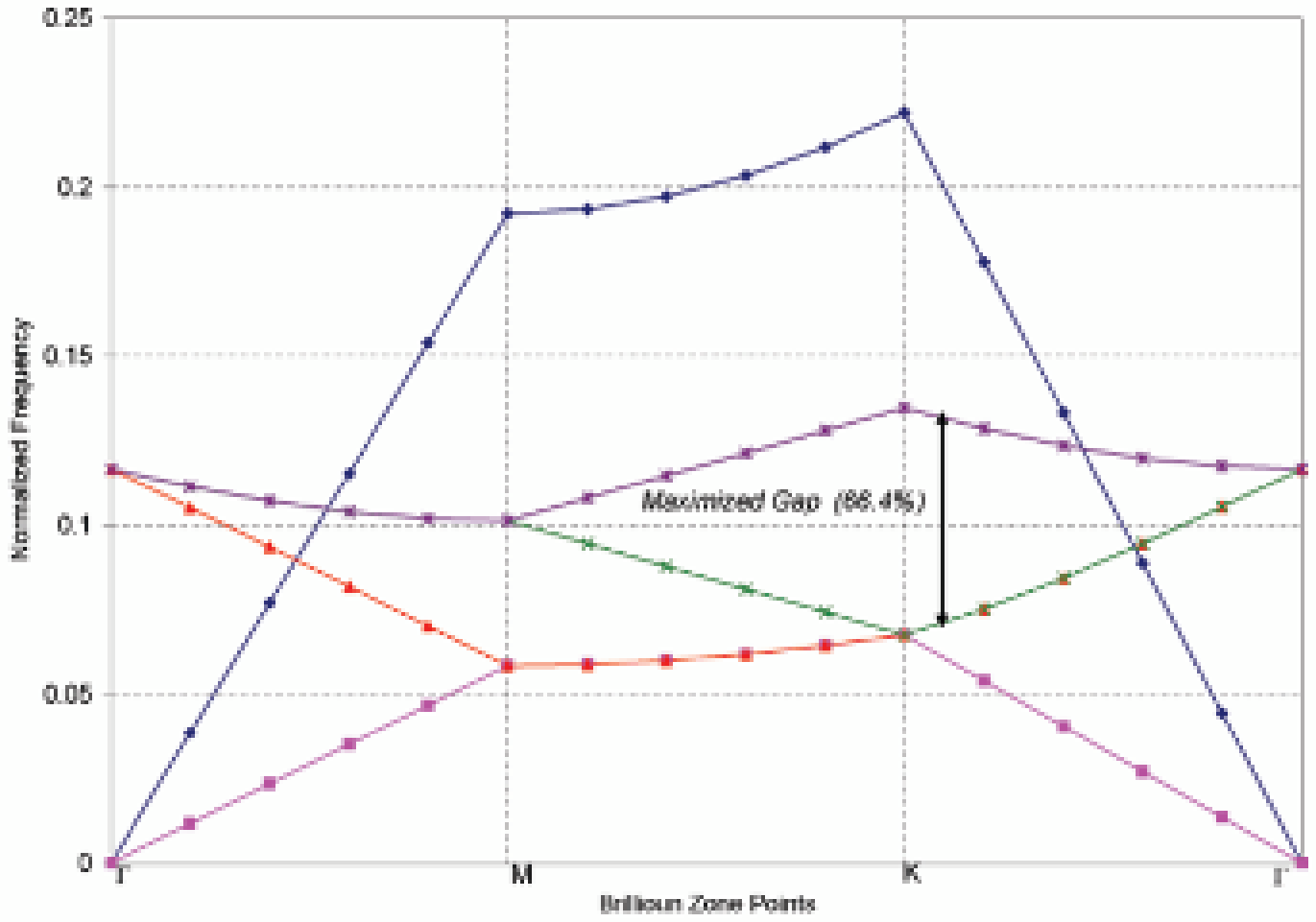}\end{overpic}}}
    \caption{Dielectric structures (\emph{a-d}), showing the optimal PC structures
              predicted by 4 runs our Genetic Algorithm. The unit cell for each structure is
              depicted by the yellow bounding box. A sample band diagram (for Run 3) is
              shown in (\emph{e}). The optimized TE-bandgap, calculated as the ratio of
              the size of the gap to the midgap value, was found to be~$\simeq$~72\%. The
              TE-bandgap for a triangular lattice with uniform air holes (r/a = 0.3) is
              shown in (\emph{f}) for reference.}
    \label{optimal-structure-run3}
\end{figure}

Our Genetic Algorithm performs as expected, and we get a general increase of fitness as
the algorithm progresses. All the four runs do not show any significant increase in
fitness after Generation 80, at which point they have maximum fitnesses (i.e. ratio of
their bandgap to midgap value) of around 72\%. All the optimized structures after the
run have similar dielectric structures and band diagrams. The dielectric structures and
a sample band diagram is shown in Figure~\ref{optimal-structure-run3}.
\medskip

\medskip

\subsection{Optimal dual PC structures}
As a more complex example, let us consider two similar PC designs, (1) a triangular
lattice of air holes in a dielectric slab, and (2) a triangular lattice of dielectric
rods in air. Structure (1) possesses a bandgap for TE light, but no bandgap for TM
light, while structure (2) possesses a bandgap for TM light, but not for TE light.

Our objective is to use the Genetic Algorithm to find a PC design in which the TE
eigenmode for structure (1) and the TM eigenmode for structure (2) are most similar.
Maxwell's equations can be cast as eigenproblems for the Electric or Magnetic fields,
and our approach could be potentially useful in future PC design, because solving the
inverse problem is analytically simpler (at least intuitively) for the eigenproblem
involving the \emph{E}-field.

\begin{figure}[htp]
    \centering
    \subfloat[Band 1, E-field]{\label{max_match_band1_e}
    \begin{overpic}[width=3.5cm]{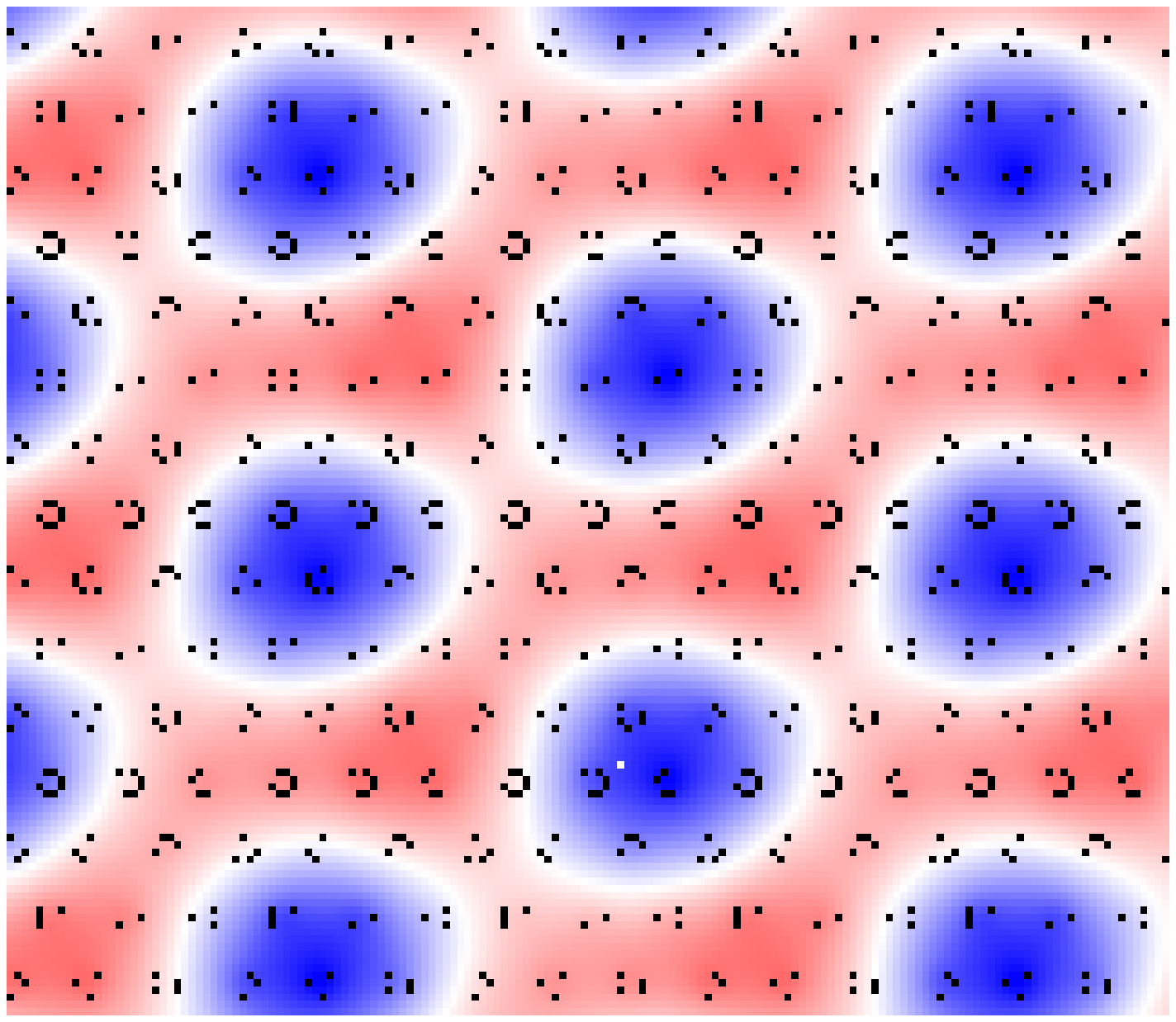}
    \end{overpic}}
    \subfloat[Band 1, H-field]{\label{max_match_band1_h}
    \begin{overpic}[width=3.5cm]{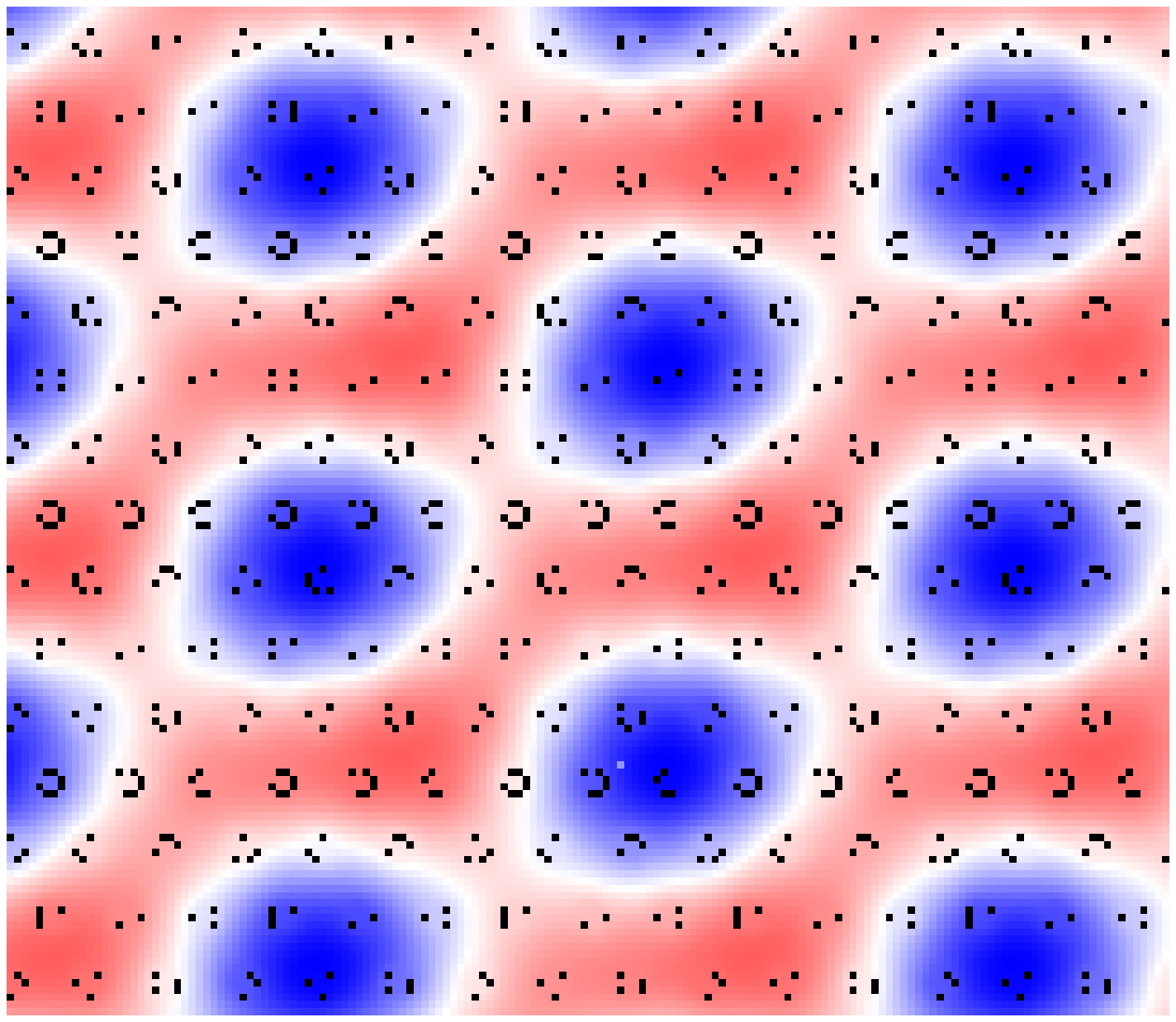}
    \end{overpic}} \\
    \subfloat[Band 2, E-field]{\label{max_match_band2_e}
    \begin{overpic}[width=3.5cm]{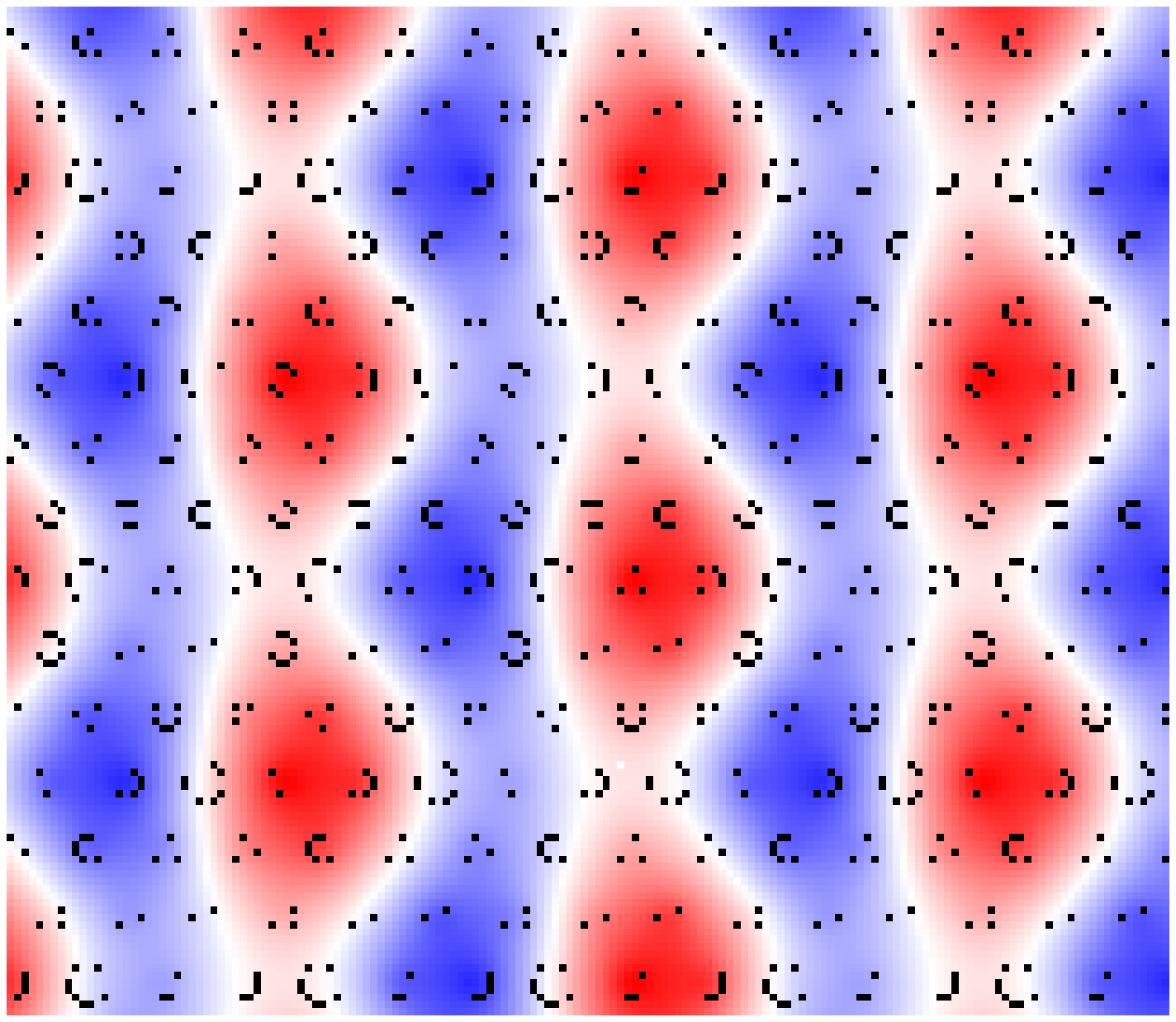}
    \end{overpic}}
    \subfloat[Band 2, H-field]{\label{max_match_band2_h}
    \begin{overpic}[width=3.5cm]{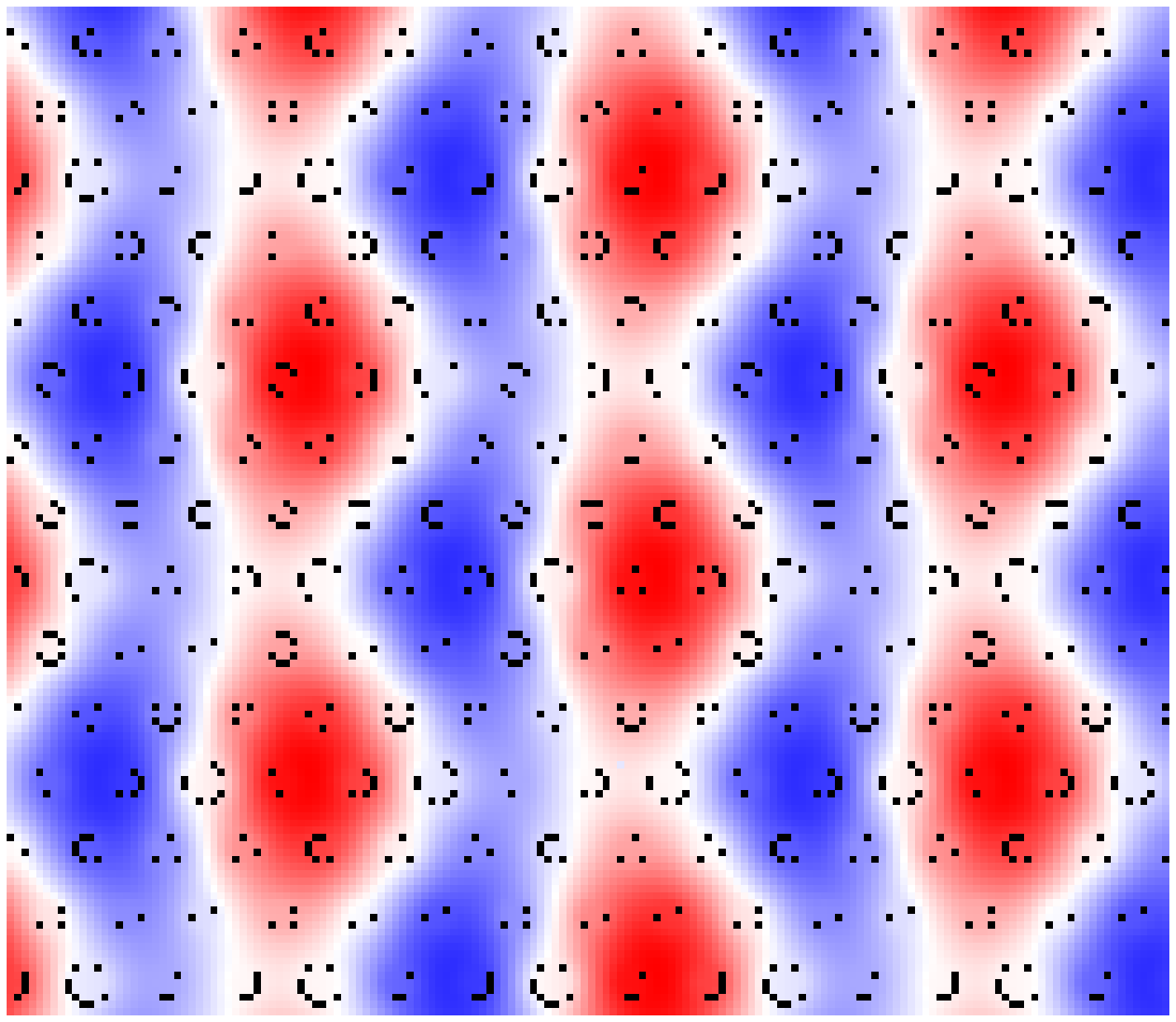}
    \end{overpic}} \\
    \subfloat[Band 3, E-field]{\label{max_match_band3_e}
    \begin{overpic}[width=3.5cm]{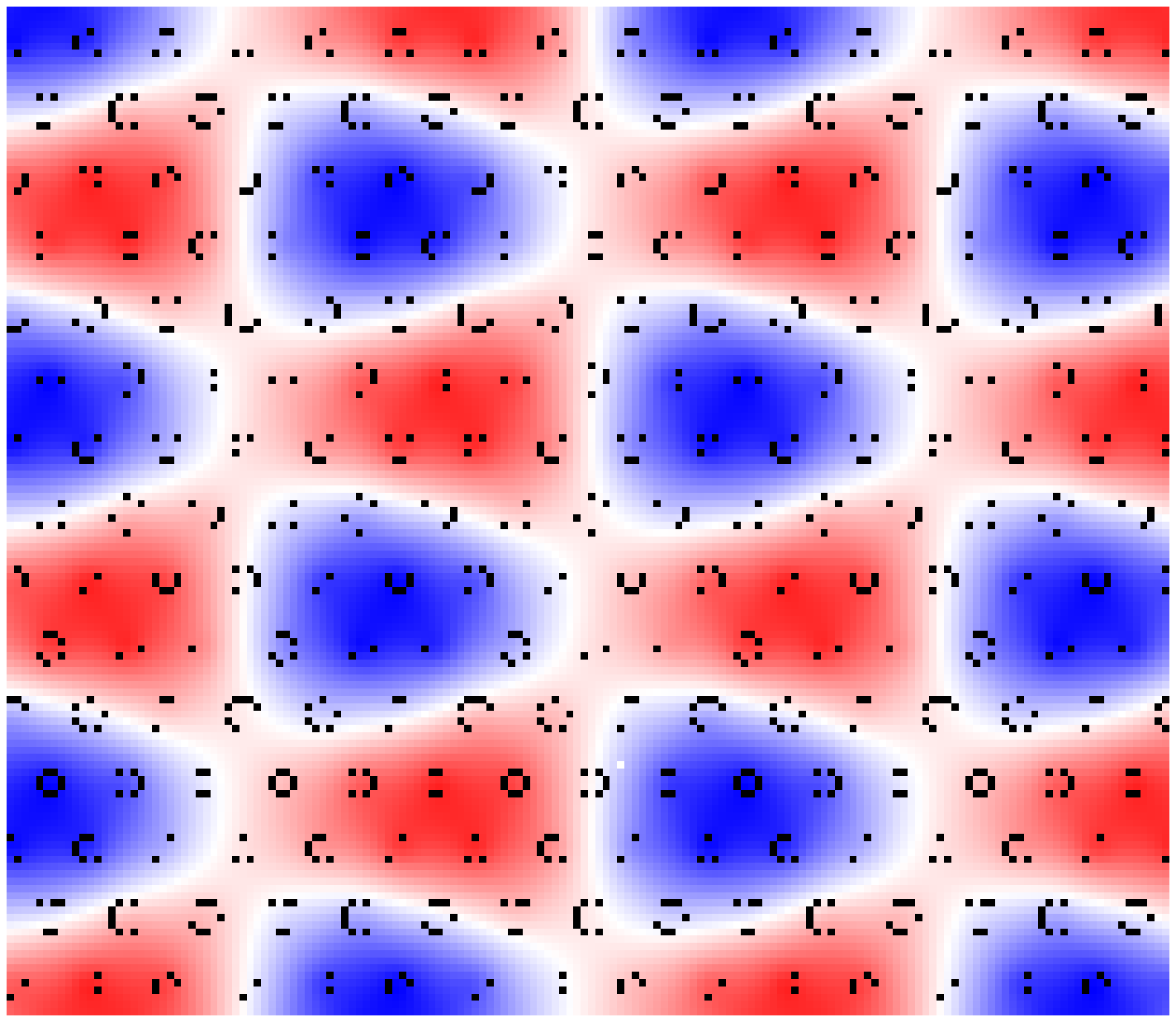}
    \end{overpic}}
    \subfloat[Band 3, H-field]{\label{max_match_band3_h}
    \begin{overpic}[width=3.5cm]{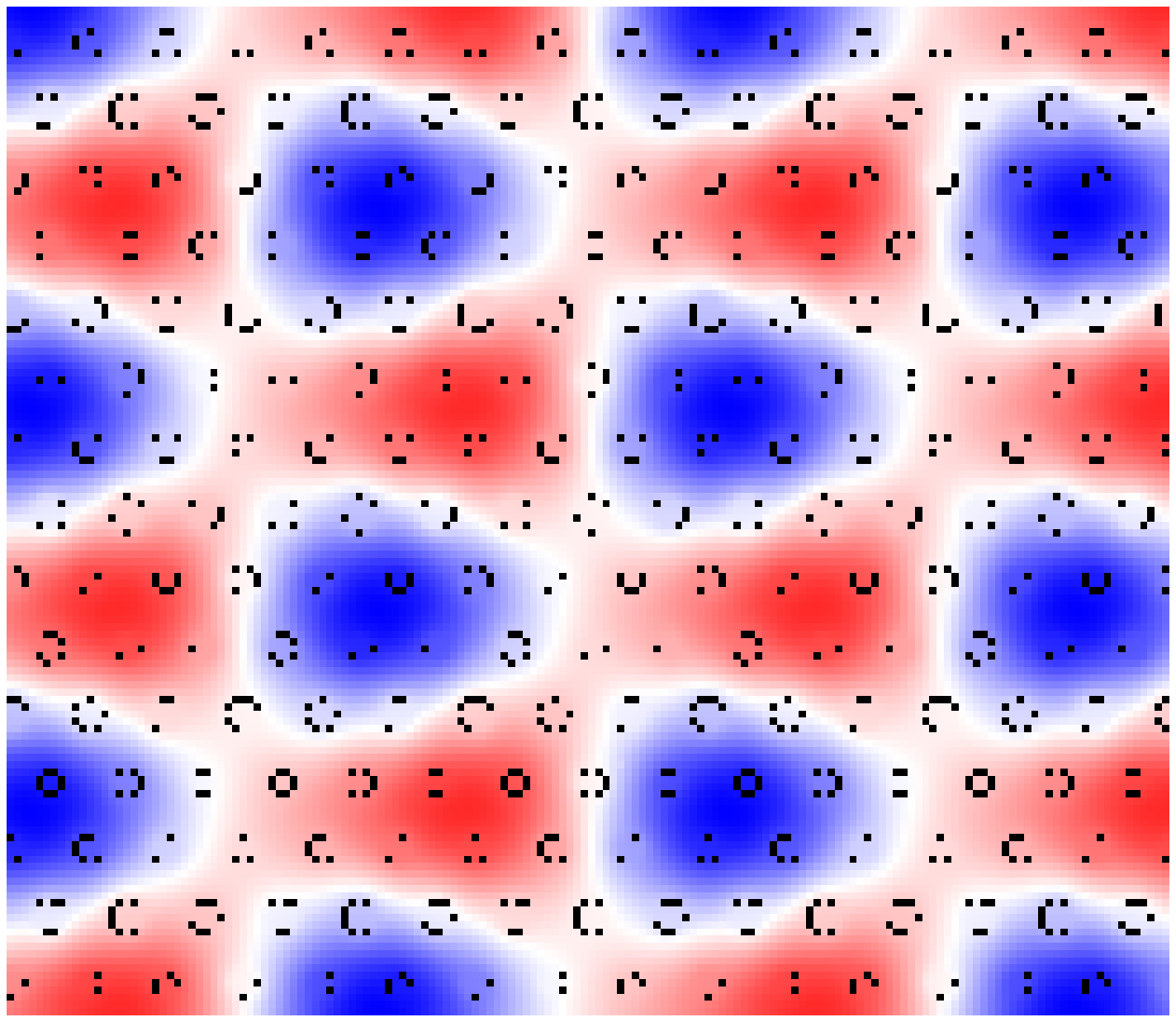}
    \end{overpic}} \\
    \subfloat[Band 4, E-field]{\label{max_match_band4_e}
    \begin{overpic}[width=3.5cm]{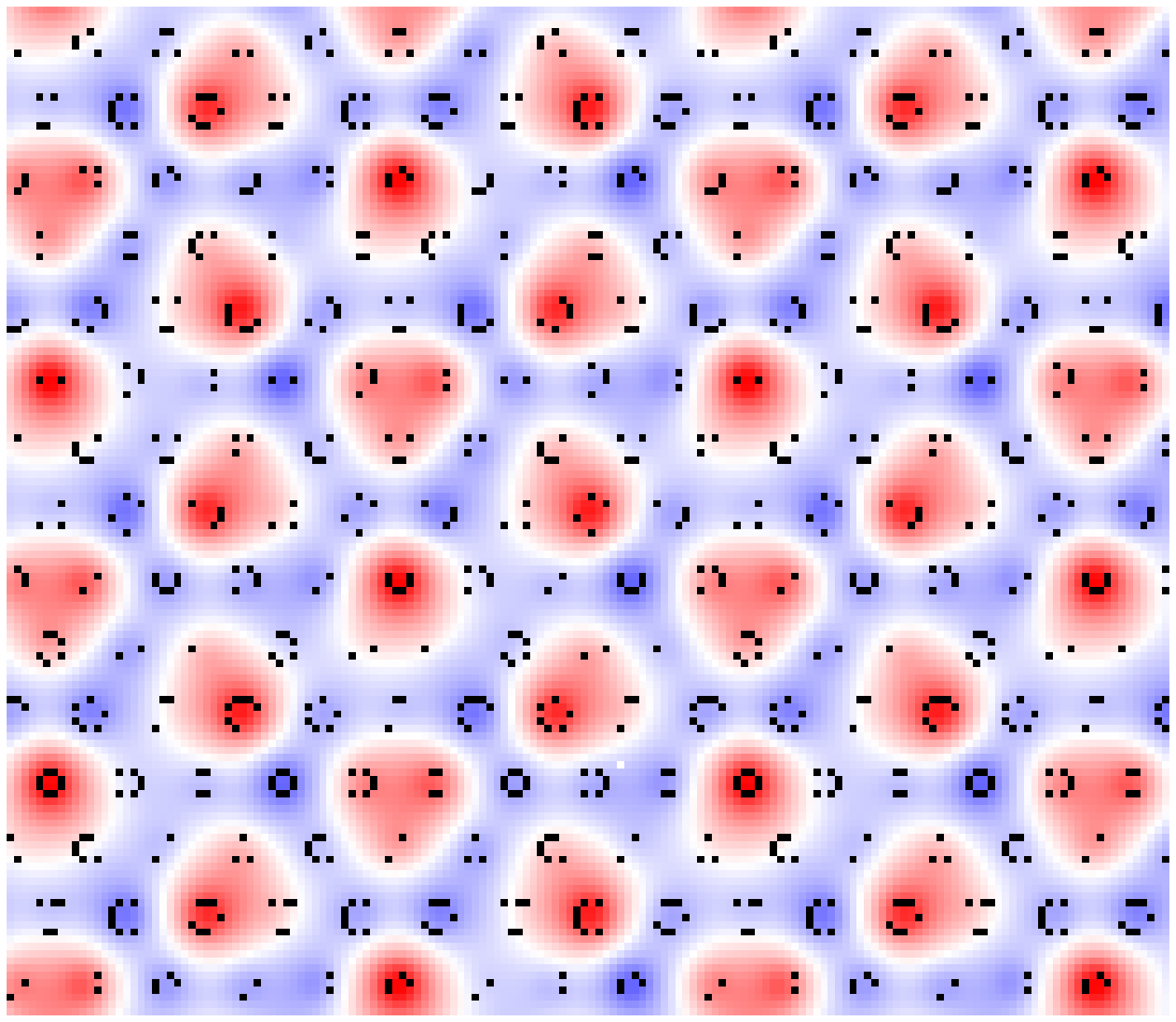}
    \end{overpic}}
    \subfloat[Band 4, H-field]{\label{max_match_band4_h}
    \begin{overpic}[width=3.5cm]{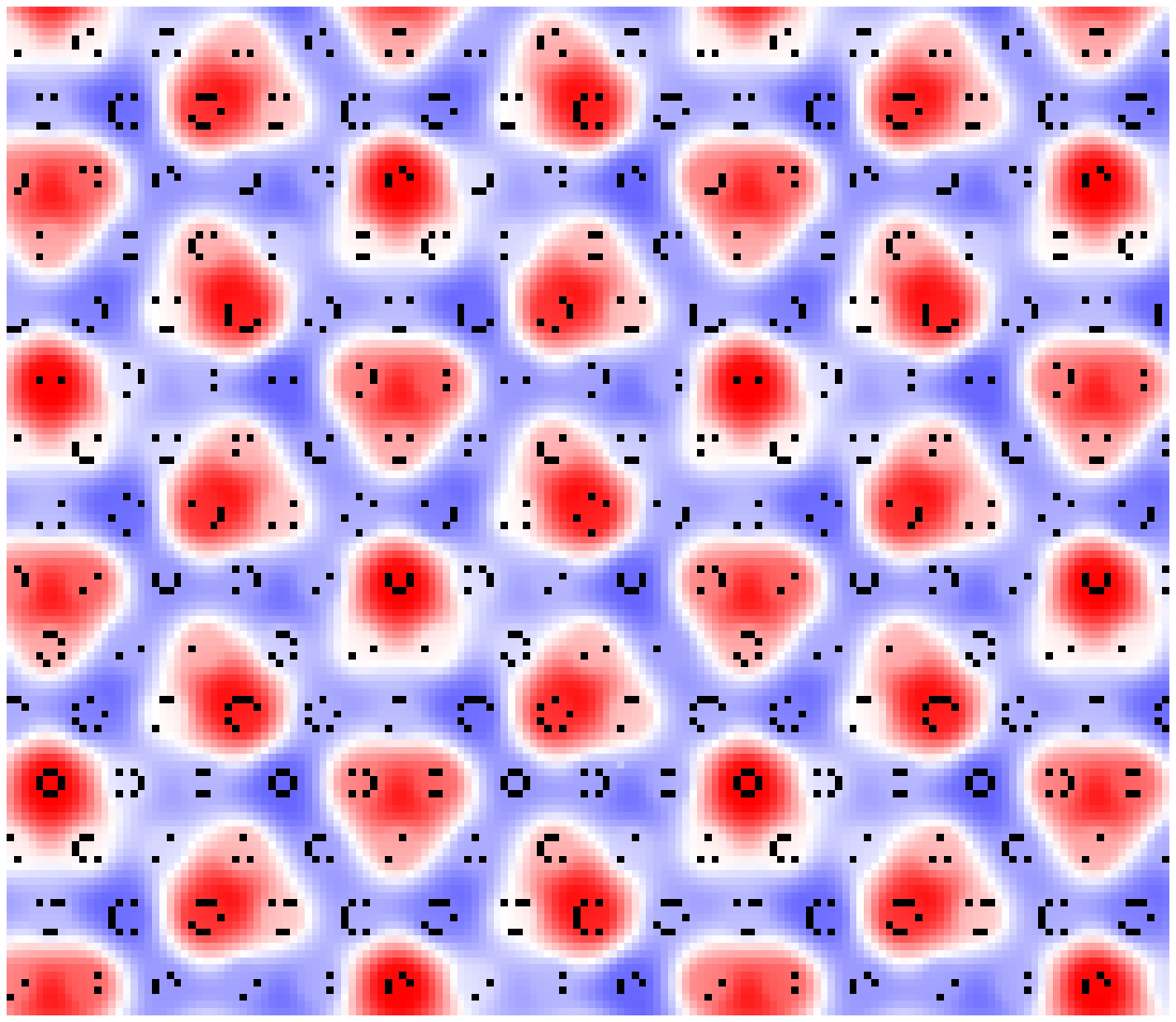}
    \end{overpic}}
    \caption{Genetic Algorithm prediction of PC structures that have optimally matched E
                and H fields, for the lowest 4 bands, at the K point. The E-fields are shown for
                structure with dielectric rods, and have a TM bandgap, while the H-fields
                are shown for structures with air holes, and have a TE bandgap.
                The shown fields are in the direction aligned with the rods.
                The fields for the lowest 3 bands are very well matched, but begin to
                deviate significantly from each other at band 4.}
    \label{max_match}
\end{figure}

We used a 3x3 supercell for the optimization, and we minimize the mean-square difference
of the z-components of the electric and magnetic fields of the dual structures at the
K-point of the band diagram. We recognize \emph{a priori} that a trivial solution, which
we wish to avoid, is a structure that has a uniform refractive index (either dielectric
or air) throughout, and so we prevent the genetic algorithm from obtaining this by
restricting our mutation to only a Gaussian mutation (see Eq.~\ref{gauss_mutation}).
This preferentially searches the locality of points, and is a necessary trade-off for
obtaining a reasonable solution. This illustrates the versatility of the Genetic
approach - the extent of the search can be easily modified by a simple change of
algorithm parameters. Fig.~\ref{max_match} shows the optimal dual structures with the
corresponding simulated fields.

\medskip

\section{Conclusion}
From the results above, we have shown that our Genetic Algorithm is able to effectively
optimize PC designs to meet specific design criteria. Furthermore, by our choice of
encoding, we could easily impose constraints upon the design space to ensure that every
design searched by the algorithm could be realistically fabricated. Between different
optimizations, all that needed to be changed was the measure of how well a given
structure complied with our design criterion - the "fitness function" in Genetic
Algorithm parlance. Our Genetic Algorithm is therefore highly robust and can be easily
modified to optimize any user-defined objective function.


\end{document}